\documentclass{aa}  

\usepackage{graphicx}
\usepackage{txfonts}
\usepackage{mathrsfs}
\usepackage[colorlinks,allcolors=blue]{hyperref}
\usepackage{breakurl}
\usepackage{subfigure}

\def \src {\mbox{A0538$-$66}}
\begin{document}

   \title{Exploring the role of X-ray reprocessing and irradiation in the anomalous bright optical outbursts of A0538$-$66}
\titlerunning{X-ray reprocessing and irradiation in the optical outbursts of A0538$-$66}

   \author{L. Ducci
          \inst{1,2}
          \and
          S. Mereghetti\inst{3}
          \and
          K. Hryniewicz\inst{4}
          \and
          A. Santangelo\inst{1}
          \and
          P. Romano\inst{5}          
          }

   \institute{Institut f\"ur Astronomie und Astrophysik, Kepler Center for Astro and Particle Physics, Eberhard Karls Universit\"at, 
              Sand 1, 72076 T\"ubingen, Germany\\
              \email{ducci@astro.uni-tuebingen.de}
              \and 
              ISDC Data Center for Astrophysics, Universit\'e de Gen\`eve, 16 chemin d'\'Ecogia, 1290 Versoix, Switzerland
              \and
              INAF -- Istituto di Astrofisica Spaziale e Fisica Cosmica, Via E. Bassini 15, 20133 Milano, Italy
              \and
              Nicolaus Copernicus Astronomical Center, Polish Academy of Sciences, Bartycka 18, 00-716 Warsaw, Poland
              \and
              INAF -- Osservatorio Astronomico di Brera, via Bianchi 46, 23807 Merate (LC), Italy
             }

   \date{Received ...; accepted ...}

 
  \abstract
   {In 1981, the Be/X-ray binary \src\ showed outbursts
    characterized  by high peak luminosities
    in the X-ray ($L_{\rm x} \approx 10^{39}$\,erg\,s$^{-1}$) and optical 
    ($L_{\rm opt} \approx 3\times 10^{38}$\,erg\,s$^{-1}$) bands.
    The bright optical outbursts were qualitatively explained as X-ray reprocessing in a gas cloud
    surrounding the binary system.}
   {Since then, further important information about the properties of \src\ have been obtained, and
    sophisticated photoionization codes have been developed to calculate the radiation
    emerging from a gas nebula illuminated by a central X-ray source.
    In the light of the new information and tools available, 
    we considered it was worth studying again the enhanced optical emission displayed by \src\
    to understand the mechanisms responsible for these unique events among the class 
    of Be/X-ray binaries.}
   {We performed about $10^5$ simulations of a gas envelope surrounding the binary system photoionized
    by an X-ray source.
    We assumed for the shape of the gas cloud either a sphere or a circumstellar disc observed edge-on.
    We studied the effects of varying the main properties of the envelope
    (shape, density, slope of the power law density profile, size) and the influence
    of different input X-ray spectra and X-ray luminosity  on the optical/UV emission
    emerging from the photoionized cloud.
    We determined the properties of the cloud and the input X-ray emission by comparing
    the computed spectra with the \emph{IUE} spectrum and photometric $UBV$ measurements obtained during 
    the outburst of 29 April 1981. 
    We also explored the role played by the X-ray heating of the surface of the donor star and the accretion
    disc irradiated by the X-ray emission of the neutron star.}
   {
     We found that reprocessing in a spherical cloud with a shallow radial density distribution
     and size of about $3 \times 10^{12}$\,cm can reproduce the optical/UV emission
     observed on 29 April 1981.
     To our knowledge, this configuration has never been observed either in \src\ during other epochs
     or in other Be/X-ray binaries.
     We found, contrary to the case of most other Be/X-ray binaries, that the optical/UV radiation produced by the X-ray heating 
     of the surface of the donor star irradiated by the neutron star is non-negligible, 
     due to the particular orbital parameters of this system 
     that bring the neutron star very close to its companion.}
   {}


   \maketitle
%

\section{Introduction} 
\label{sect intro}
Be/X-ray binaries (Be/XRBs), first recognized as an independent class 
of X-ray sources by \citet{Maraschi76},
are the most conspicuous members of high-mass X-ray binaries  (HMXBs; for a review, see e.g. \citealt{Reig11}).
They consist of a Be star and, usually, a neutron star (NS).
Most of them show a weak persistent X-ray emission sporadically interrupted by outbursts
lasting from a few to several weeks. The X-ray outbursts are caused by the accretion onto the NS
of the matter captured from the circumstellar disc around the Be star.
Optical properties of Be/XRBs have been studied since their discovery (e.g. \citealt{Tarenghi81}),
 recently by using the extensive photometric observations of the Magellanic Clouds obtained
by the MACHO and OGLE projects (e.g. \citealt{Rajoelimanana11, Bird12, Schmidtke13}).
The optical emission of Be/XRBs is dominated by the radiation emitted
at the surface of the Be star and by the circumstellar disc.
Some Be/XRBs show long-term ($200- \gtrsim 3000$\,d) irregular
or quasi-periodic variations, with amplitudes $\Delta m_{\rm I} \lesssim 0.6$ \citep{Rajoelimanana11}.
This variability is likely linked to the formation and depletion 
of the circumstellar disc. The pattern of variability
changes from source to source, and also depends  on the inclination angle of the circumstellar
disc with respect to the line of sight (see e.g. \citealt{Haubois12}).
A small group of Be/XRBs also shows  orbital periodicities in the optical band
with lower amplitudes ($\Delta m_{\rm I} \approx 0.01-0.2$).
This type of variability is usually attributed to the 
circumstellar disc perturbed by the orbiting NS.
Unfortunately, a detailed model to explain the optical variability 
on orbital timescales is still missing.

\begin{figure*}
\begin{center}
\includegraphics[width=\columnwidth+\columnwidth]{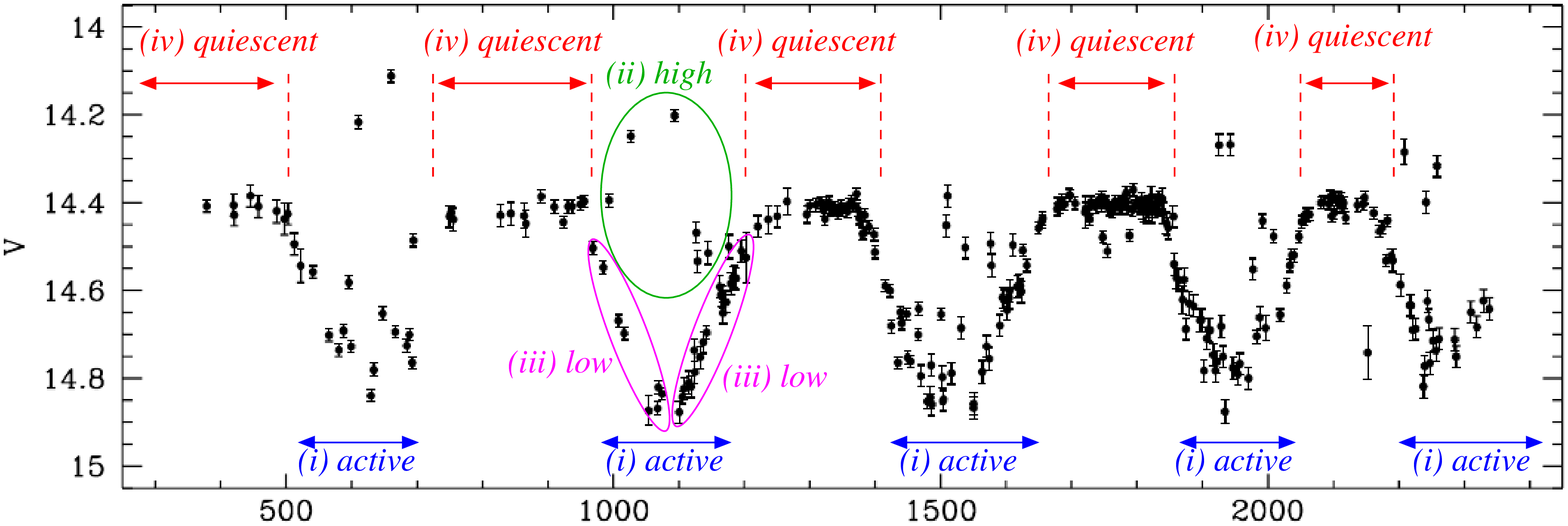}\\
{\footnotesize JD $-$ 2448623.5}
\end{center}
\caption{\footnotesize V-band light curve of \src\
          from MACHO project observations (figure adapted from figure 1 in \citealt{Alcock01}).
         The quiescent, active states (low and high)  of the \citet{McGowan03} 
         and \citet{Alcock01} model are shown.}
\label{figalcock}
\end{figure*}

A remarkable exception to the typical low-amplitude optical variability observed
in Be/XRBs is 1A\,0538$-$66 (hereafter \src), 
which  in 1981 showed recurrent optical outbursts
with an increase of the optical brightness of $\approx$2.3 magnitudes \citep{Charles83}.
\src, located in the Large Magellanic Cloud (LMC),
hosts a NS with a pulse period of 69~ms 
and orbital period of $\sim 16.6$~d \citep{Skinner82},
which corresponds to the recurrence period of the optical outbursts.
It was discovered in 1977 \citep{White78} 
during an X-ray outburst that reached
a peak luminosity of $8.5\times 10^{38}$~erg~s$^{-1}$.
Other super-Eddington outbursts were subsequently observed.
\emph{ROSAT}, \emph{BeppoSAX}, \emph{ASCA}, and \emph{XMM-Newton} \citep{Campana02, Kretschmar04}
revealed outbursts with lower luminosities 
($L_x \approx 10^{33} - 10^{37}$\,erg\,s$^{-1}$).
The outbursts of \src\ show a wide range of durations, from
a few hours to $\sim 14$~days,
and a high dynamical range spanning five
orders of magnitude. 
 As mentioned above, \src\ displays a unique optical behaviour when compared with other Be/X-ray transients.
\citet{Skinner80,Skinner81} discovered bright and recurrent
optical outbursts (up to $L_{\rm opt} \approx 3\times 10^{38}$\,erg\,s$^{-1}$; \citealt{Charles83})
with period of $\sim 16.6$\,d in phase with the X-ray outbursts.
A photometric light curve from 1915 to 1998,
based on MACHO data and UK Schmidt and Harvard photographic
B-band plates revealed, in addition to the recurrent outbursts
with period $\sim 16.6$\,d, a long-term modulation at $P_{\rm sup}=420.8 \pm 0.8$
days, with a reddening at low fluxes, in agreement with the formation
of a circumstellar disc \citep{Alcock01,McGowan03}.
\citet{Alcock01} and \citet{McGowan03} suggested that the long-term modulation is caused
by the depletion and formation of a circumstellar disc
around the Be star, observed nearly edge-on.
The orbital modulation is seen only at certain phases of the 421-day cycle,
suggesting that $P_{\rm sup}$ is caused by variations in the Be star envelope.
When the NS is embedded in the high-density wind of the circumstellar disc 
(see \emph{active} state in Fig. \ref{figalcock}),  
it accretes material which leads to X-ray and optical outbursts 
(see \emph{high} state in Fig. \ref{figalcock}).
This scenario is consistent with the observation of Balmer emission lines  
during the active state, testifying the presence of a circumstellar disc.  
When the optical flux is high for a long period (the \emph{quiescent} state, 
see Fig. \ref{figalcock}), the optical/UV spectrum does not show any 
Balmer emission, suggesting the absence of the circumstellar disc.  
Therefore, during the quiescent state, only the naked B star is observed.
The exceptionally bright optical outbursts of \src\ observed until about 1981, 
were relatively long ($t\approx 3$ days, see figure 1 in \citealt{Densham83}).
Subsequently, the source showed a different type of outbursts, faster ($t\lesssim 1$\,d)
and fainter ($\Delta m \approx 0.5$),
associated with X-ray outbursts with lower luminosity 
($L_{\rm x}<10^{37}$\,erg\,s$^{-1}$; \citealt{Corbet97, Alcock01}).
Recently, it has been shown that \src\ still shows fast optical outbursts 
\citep{Ducci16, Rajo17}.

\citet{Densham83}, \citet{Maraschi83}, and \citet{Howarth84}, proposed a qualitative scenario
to explain the bright optical outbursts observed in 1980 and 1981.
The scenario requires that a dense gas cloud forms around the binary system
from the material tidally displaced by the NS from the outer layers of the donor star
over many orbits.
The bright optical outbursts are then powered by the X-ray photons produced by the accreting 
NS which are reprocessed in the envelope.
\citet{Maraschi83} considered the models of high-density clouds illuminated by a central X-ray source
proposed by  \citet{Hatchett76} and \citet{Kallman82}.
Scaling the model calculations presented by these authors, \citet{Maraschi83}
found that the effective temperature of the optical radiation emitted during the optical
outburst ($T\approx10^4$\,K) could be obtained assuming an X-ray luminosity of $L_{\rm x}=10^{39}$\,erg\,s$^{-1}$
and a homogeneous sphere of gas with density $n=10^{12}$\,cm$^{-3}$ and size of $R_{\rm out}=3\times10^{12}$\,cm.
\citet{Apparao88} used a similar approach to show that the enhanced continuum and emission lines
of the optical outbursts are produced by a disc of gas around the Be star ionized by the
X-ray radiation emitted by the NS during the periastron passage, when it crosses the circumstellar disc.
\citet{Apparao88} used a model developed by \citet{Kallman82}, which consists of an X-ray 
source with a 10\,keV bremsstrahlung spectrum that ionizes a sphere with uniform
density surrounding it. On the basis of this model, \citet{Apparao88} considered the ionized 
states of H, He, and \ion{C}{IV} to calculate the continuum emission produced by an irradiated
disc with height $5\times 10^{11}$\,cm, a uniform density, and a given size.
They calculated the continuum emission for different values of the input X-ray luminosity,
density, and disc radius, 
assuming that the distance of the NS from the Be star at periastron is $5\times 10^{12}$\,cm.
\citet{Apparao88} found the best agreement between the computed emission 
and the observed ultraviolet and optical continuum of \src\ during an outburst
for $L_{\rm x} = 2.5 \times 10^{39}$\,erg\,s$^{-1}$,
$n=2\times 10^{11}$\,cm$^{-3}$, and $R_{\rm out}=1.38\times 10^{13}$\,cm.

The calculations presented by \citet{Maraschi83} and \citet{Apparao88} were based on a simplified general
class of models computed by \citet{Kallman82} and \citet{Hatchett76} proposed for a wide variety 
of astrophysical objects. Although they could be rescaled, they were not tailored
for the specific case of \src.
In particular, we refer to the input X-ray spectrum, the geometry of the system,
and the density properties of the nebula and the abundances, which were all fixed in the 
\citet{Kallman82} model to values not suitable for the case of \src.
Therefore, we used the photoionization code CLOUDY (v17.01) to compute the emission
spectrum of a cloud ionized by the X-ray radiation produced by the accreting pulsar of \src. 
CLOUDY is designed to simulate the physical conditions and emission spectra
of a photoionized gas. It  can be used to describe a wide variety of
astrophysical situations \citep{Ferland17}.
The code requires as input the spectrum and luminosity of the X-ray source,
the geometry of the cloud (we assume different sizes,
shapes, and density profiles) and gas composition (we assume the typical
metallicity of the LMC). 
The ionization and temperature structure are found by CLOUDY assuming
a local balance between heating and cooling and between ionization
and recombination.
The processes taken into account include 
photoionization, Auger effect, Compton heating, charge transfer, 
collisional de-excitation, radiative and dielectronic recombination,
bremsstrahlung, collisional ionization, and collisional excitation
of bound levels.
The radiation field is determined by solving the radiative transfer
equation for both the continuum and line components (for further 
details see the documentation of CLOUDY,\footnote{\url{http://www.nublado.org/}}
and e.g. \citealt{Ferland17}).

\section{Calculations}
\label{sect. calculations}

We assumed that \src\ consists of a B1\,III star with mass $M_{\rm d}=9$\,M$_\odot$
and radius $R_{\rm d}=10$\,R$_\odot$ and a NS with mass $M_{\rm NS}=1.44$\,M$_\odot$. The 
$M_{\rm d}=9$\,M$_\odot$ value corresponds to the upper limit for the mass of the donor star
derived by \citet{Rajo17} from the mass function assuming $M_{\rm NS}=1.44$\,M$_\odot$
and from the absence of an X-ray eclipse.
Because of the lack of information needed to determine the mass function of the X-ray
pulsar, the two masses cannot be measured separately.
Therefore, the values we assumed for $M_{\rm d}$ and $M_{\rm NS}$ may slightly deviate
from the real values of the system.
For the orbital parameters we assumed $P_{\rm orb}=16.6409$\,d and eccentricity $e=0.72$ \citep{Rajo17}.

\begin{figure}
    \centering
    \includegraphics[width=\columnwidth]{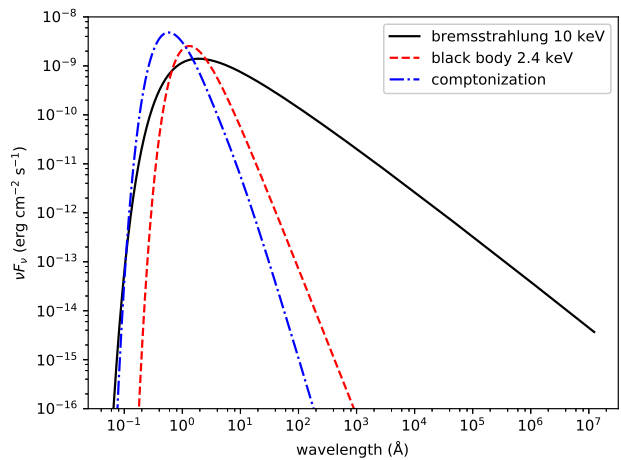}
       \caption{Comparison between the input spectra used to compute the
       spectrum emerging from the gas cloud illuminated by the NS and the donor star.
       The luminosity in the energy range 1$-$20\,keV is the same for all the input
       spectra ($L_{\rm x}=10^{39}$\,erg\,s$^{-1}$).}
          \label{confrspettriX}
    \end{figure}

Simulations are divided into three main groups. In each group, we considered a different
input spectrum:
\begin{itemize}
\item bremsstrahlung, with temperature $kT=10$\,keV;
\item black body, with temperature $kT=2.4$\,keV;
\item Comptonization of soft photons in a hot plasma ({\sc comptt} in Xspec;
\citealt{Titarchuk94, Arnaud96}), with input soft photon (Wien law) temperature $kT_0=1.35$\,keV,
plasma temperature $kT=5.5$\,keV, and  plasma optical depth $\tau = 17$ (based on the
spectrum of V0332+53 during an outburst of $L_x\approx 10^{38}$\,erg\,s$^{-1}$; 
see \citealt{Doroshenko17}).
\end{itemize}
For the first group, we considered a bremsstrahlung model with temperature $kT=10$\,keV
to allow us to compare our results with those obtained by \citet{Apparao88},
although the results from this model should be taken with caution
because of the large fraction of soft X-ray and UV photons it produces
compared to more physical models (see Fig. \ref{confrspettriX}).
For the second group of simulations, we considered the model 
found by \citet{Ponman84} that best fits the X-ray emission from \src\ during
a bright outburst observed by the \emph{Einstein} satellite.
For the third group, we considered a physical model with properties
similar to those observed in other Be/XRBs observed during high-luminosity states.
For each of these groups, we considered two geometries
for the envelope: a sphere or a disc with uniform height $h=5 \times 10^{11}$\,cm
(similarly to \citealt{Apparao88}).

We considered for our calculations the  closed geometry used
in CLOUDY because we expect that the covering factor of the cloud
seen by the X-ray source is large  for the spherical
and for the disc case.
On the basis of previous findings,
we assumed that the circumstellar disc is observed edge-on.
We adopted for all the simulations a metallicity of $Z=0.4$\,Z$_\odot$,
according to the average values of LMC reported in \citet{Zhukovska13} and \citet{Russell92}.
The ionization parameter inside the cloud (according to the definition of \citealt{Tarter69})
in our simulations is $\lesssim 10^6$\,erg\,cm\,s$^{-1}$.

Optical and near-infrared studies showed that the disc density of all
Be stars (isolated or in Be/XRBs) follows a power-law density distribution,
\begin{equation} \label{density profile}
n(r) = n_0 \left ( \frac{r}{R_{\rm d}} \right )^{-\alpha} \mbox{ cm}^{-3} \mbox{ ,}
\end{equation}
where $n_0$ is the number density of atoms at $R_{\rm d}$
and $\alpha$ is the density slope \citep{Waters89}.
We used Eq. \ref{density profile} to describe the density distribution
also in the case of clouds with spherical symmetry.

For each group and geometry (bremmstrahlung, black body, Comptonization; sphere or disc) 
we ran $10^3$ simulations varying $n_0$,
the size of the cloud ($R_{\rm out}$), and $\alpha$ in the ranges
$10^{11} \leq n_0 \leq 10^{13}$\,cm$^{-3}$,
$2\times 10^{12} \leq R_{\rm out} \leq 3\times 10^{13}$\,cm,
$0\leq \alpha \leq 2.5$ in a grid $10\times 10 \times 10$.
The assumed values of $R_{\rm out}$ and $n_0$ were uniformly distributed
in a $\log_{10}$ space in the ranges defined above,
while $\alpha$ values were uniformly distributed in a linear space.
Although the X-ray luminosity $L_{\rm x}$ at the peak of the outbursts is reasonably well constrained
from the observations \citep{Ponman84},
\src\ was not observed in X-rays during the outburst of 29 April 1981.
Therefore, similarly to \citet{Apparao88}, we  also left   this parameter free
to vary in the range $10^{38} \leq L_{\rm x} \leq 5 \times 10^{39}$\,erg\,s$^{-1}$.

As mentioned above, we compared each spectrum computed with CLOUDY with the \emph{IUE} spectrum
and \emph{UBV} magnitudes of the outburst of 29 April 1981
reported in \citet{Charles83}.
The intensities of all spectral lines 
provided by CLOUDY are calculated correctly, although the default code does
not compute spectral line profiles\footnote{\url{https://www.nublado.org/wiki/FaqPage}}.
In addition, CLOUDY does not take into account the wind velocity;
therefore, effects like broadening of the line width or wavelength shift caused by the 
outflowing motion of the gas in the wind are not reproduced in the computed spectrum.
We thus decided to compare the computed and observed continuum in the region of
the spectrum above $2000$\,\AA, which has few prominent emission lines. 
Then we compared the luminosities of the emission lines \ion{C}{IV}\,$\lambda$1550
and \ion{He}{II}\,$\lambda$1640 computed by CLOUDY with the values reported in \citet{Apparao88}:
$L_{{\rm \ion{C}{IV}\,}\lambda1550}=3\times 10^{36}$\,erg\,s$^{-1}$;
$L_{{\rm \ion{He}{II}\,}\lambda1640}=8\times 10^{36}$\,erg\,s$^{-1}$.
For the comparison between the computed and observed continuum, we used the $\chi^2$ test.
The $\chi^2$ value is computed at each grid point for all the parameters
($\alpha$, $n_0$, $R_{\rm d}$, $L_{\rm x}$). Then the global minimum over the grid
is found at the grid point with the lowest value. The corresponding $\alpha$, $n_0$, $R_{\rm d}$, and $L_{\rm x}$,
are assumed to be the best solutions. We do not use interpolation between 
grid points in the process.

 %

\begin{table*}
\begin{center}
\caption{Best fit results and the relative reduced chi-square values ($\mathbf{\chi^2_\nu}$)
for the comparison between the optical/UV spectra
computed with {\sc CLOUDY} and the \emph{IUE}+UBV spectrum of the outburst of 29 April 1981.}
\label{Table results1}
\begin{tabular}{lcccccc}
\hline
\hline
\noalign{\smallskip}
Input spectrum  &         $R_{\rm out}$         &          $n_0$         &   $\alpha$       &  $\chi^2_\nu$  &  \ion{C}{IV}\,$\lambda$1550  & \ion{He}{II}\,$\lambda$1640  \\
                &     $\log_{10}($[cm]$)$      &   $\log_{10}($[cm]$)$   &                  &   (13 d.o.f.) & $\log_{10}$(erg\,s$^{-1}$) & $\log_{10}$(erg\,s$^{-1}$)   \\
\noalign{\smallskip}
\hline
\noalign{\smallskip}
Bremmstrahlung  &     $12.42 \pm 0.05$       &    $12.78 \pm 0.11$   &  $0.60 \pm 0.14$   &     2.12     &        35.8          &       36.1         \\
\noalign{\smallskip}
\hline
\noalign{\smallskip}
Black body       &     $12.42 \pm 0.05$       &    $12.78 \pm 0.11$   &  $0.60 \pm 0.14$  &    1.32    &        36.2            &       35.9        \\
\noalign{\smallskip}
\hline
\noalign{\smallskip}
Comptonization  &     $12.63 \pm 0.07$       &   $12.78 \pm 0.11$    &  $1.11 \pm 0.14$  &    1.44      &         36.2          &        36.2        \\
\noalign{\smallskip}
\hline
\end{tabular}
\end{center}
 Notes. The best fit parameters were obtained assuming $L_{\rm x}=2.5\times 10^{38}$\,erg\,s$^{-1}$ (1--20\,keV)
        and a spherical geometry. 
\end{table*}

\section{Results and discussion}
\label{sect. results}

\subsection{Optical/UV emission from a gas cloud ionized by the X-rays from the NS}
\label{reprocessing}

We first consider the bright outbursts of \src\ under the assumption that
they are caused by reprocessing of X-rays radiation in
the envelope surrounding the binary system.
When we assumed a circumstellar disc for the gas cloud,
we found that the luminosities of the lines \ion{C}{IV}\,$\lambda$1550 
and \ion{He}{II}\,$\lambda$1640 were 100-1000 times fainter than those observed,
and therefore this geometry was discarded.
Instead, for a spherical shell we found more acceptable values.
The difference between the luminosities of
these lines in the spherical and disc cases
 probably arises because   a sphere absorbs more energy from the source than a disc, 
and thus can reradiate it more in the form of lines.
Therefore, we list in Table \ref{Table results1}
only the best fit solutions obtained for a spherical gas shell.
Errors in Table \ref{Table results1} are at the 1\,$\sigma$ confidence level
and are calculated for each parameter independently according to the method described,
for example, in \citet{Avni76}. 
Uncertainties in Table \ref{Table results1} are comparable 
with the grid cell size,
and as such should be taken with caution.
Although the reduced $\chi^2$ value ($\chi^2_\nu$)
for the bremsstrahlung case is greater than two, the null hypothesis probability
is 0.0104. Hence, for this case, the data and model are consistent, though only at the $\approx$99\% confidence level.
For the black body  case there are two additional solutions with $\chi^2_\nu<2$
(but worse than the best fit solution shown in Table 1) for   two sets
of parameters: $\alpha=1.11\pm 0.14$, $n_0=12.78 \pm 0.11$, $R_{\rm out}=12.63 \pm 0.07$
($\chi^2_\nu=1.33$, 13 d.o.f.);
$\alpha=1.67\pm 0.14$, $n_0=13.00 \pm 0.11$, $R_{\rm out}=13.34 \pm 0.08$
($\chi^2_\nu=1.44$, 13 d.o.f.; errors  at 1$\sigma$ confidence level). 
In both cases, $L_{\rm x}=2.5\times 10^{38}$\,erg\,s$^{-1}$ (1--20\,keV).

\begin{figure*}
   \centering
   \includegraphics[width=6cm]{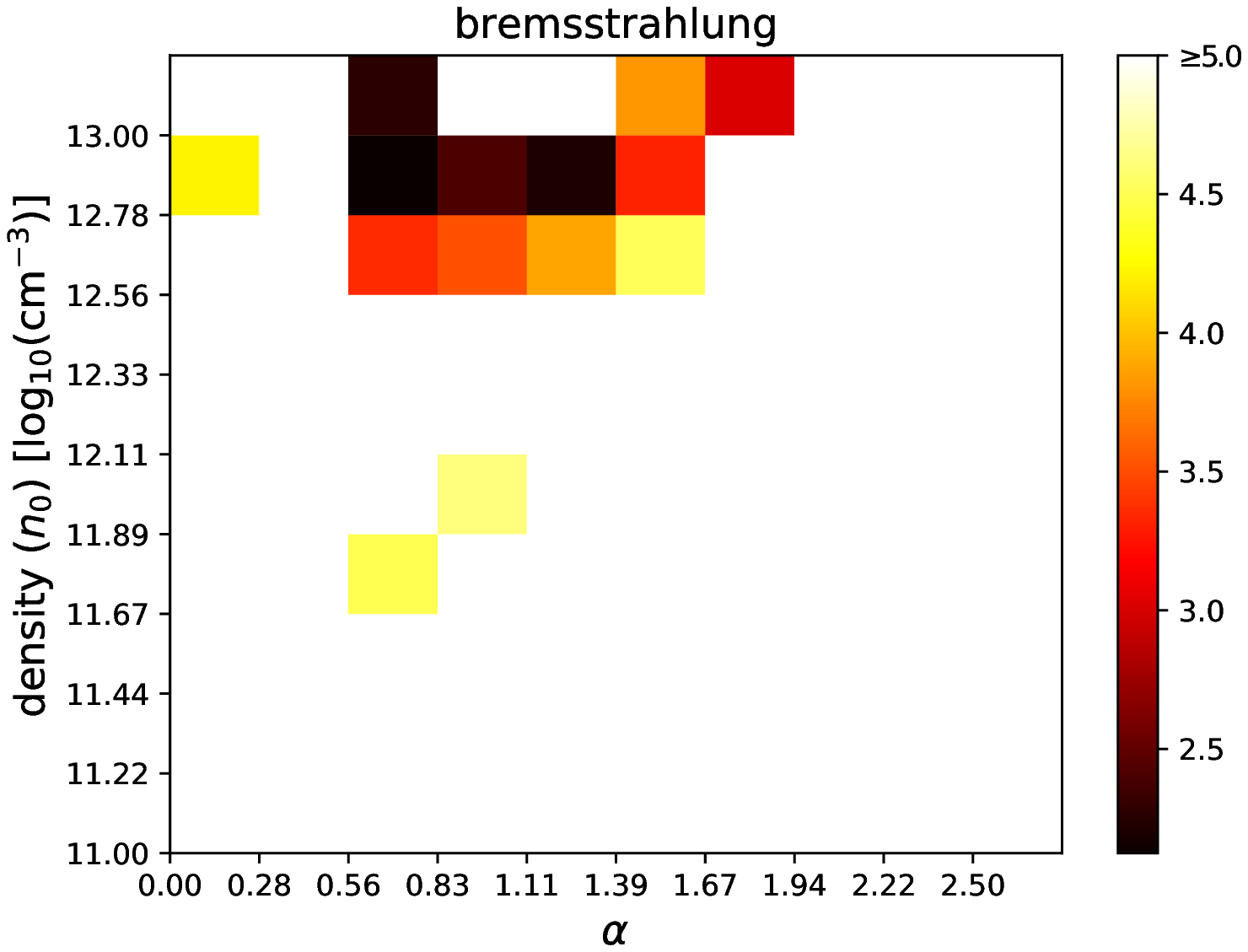}
   \includegraphics[width=6cm]{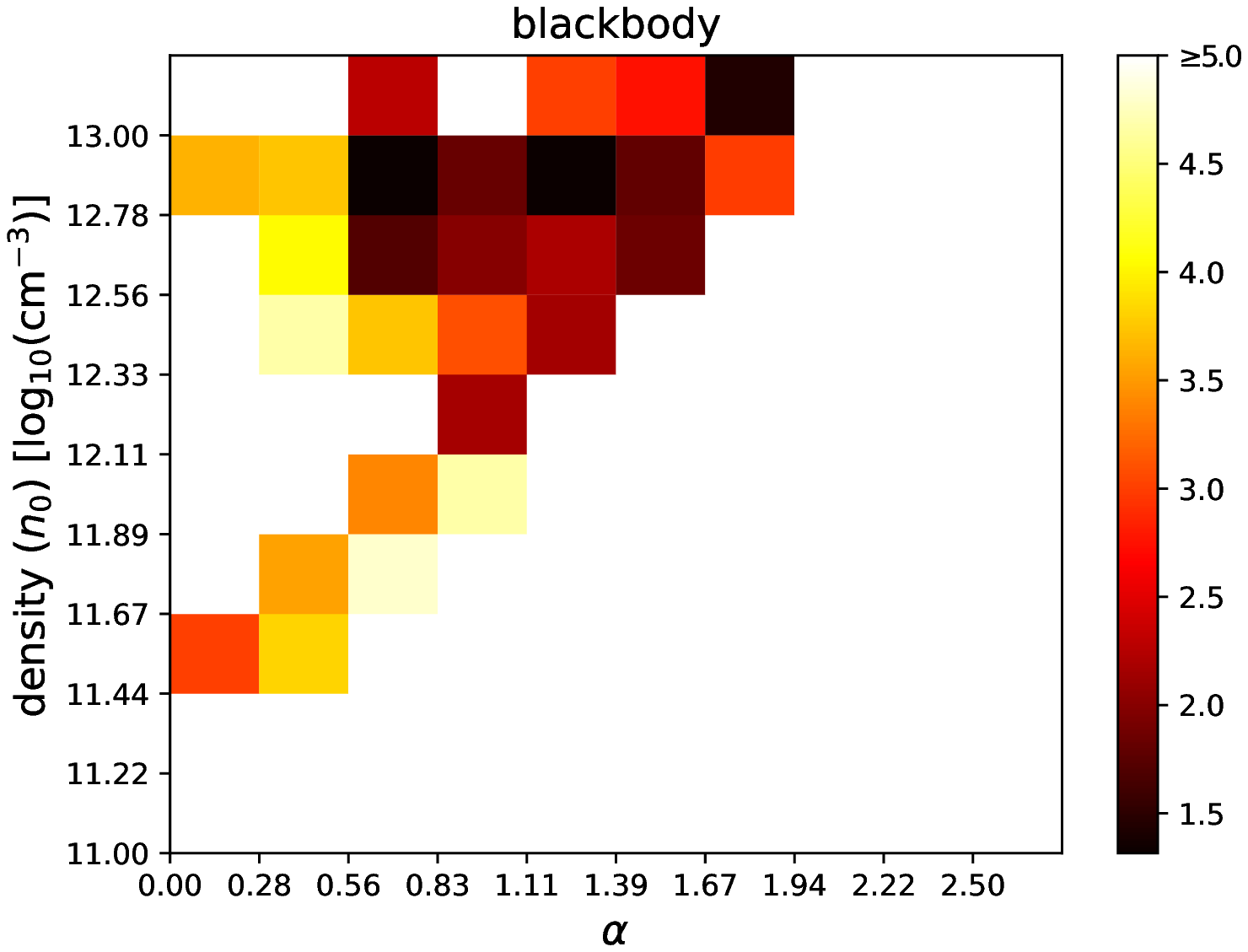}
   \includegraphics[width=6cm]{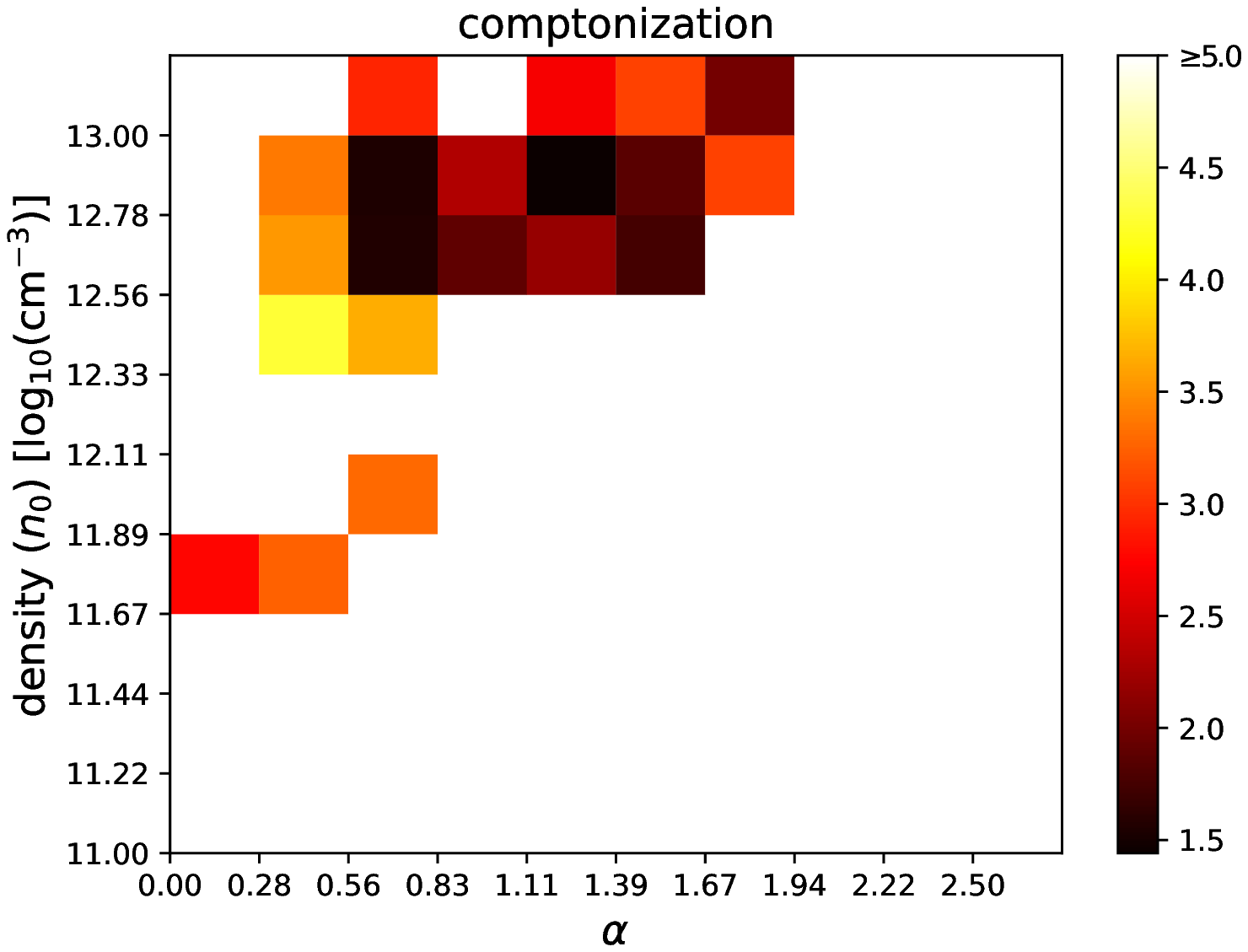}\\

   \includegraphics[width=6cm]{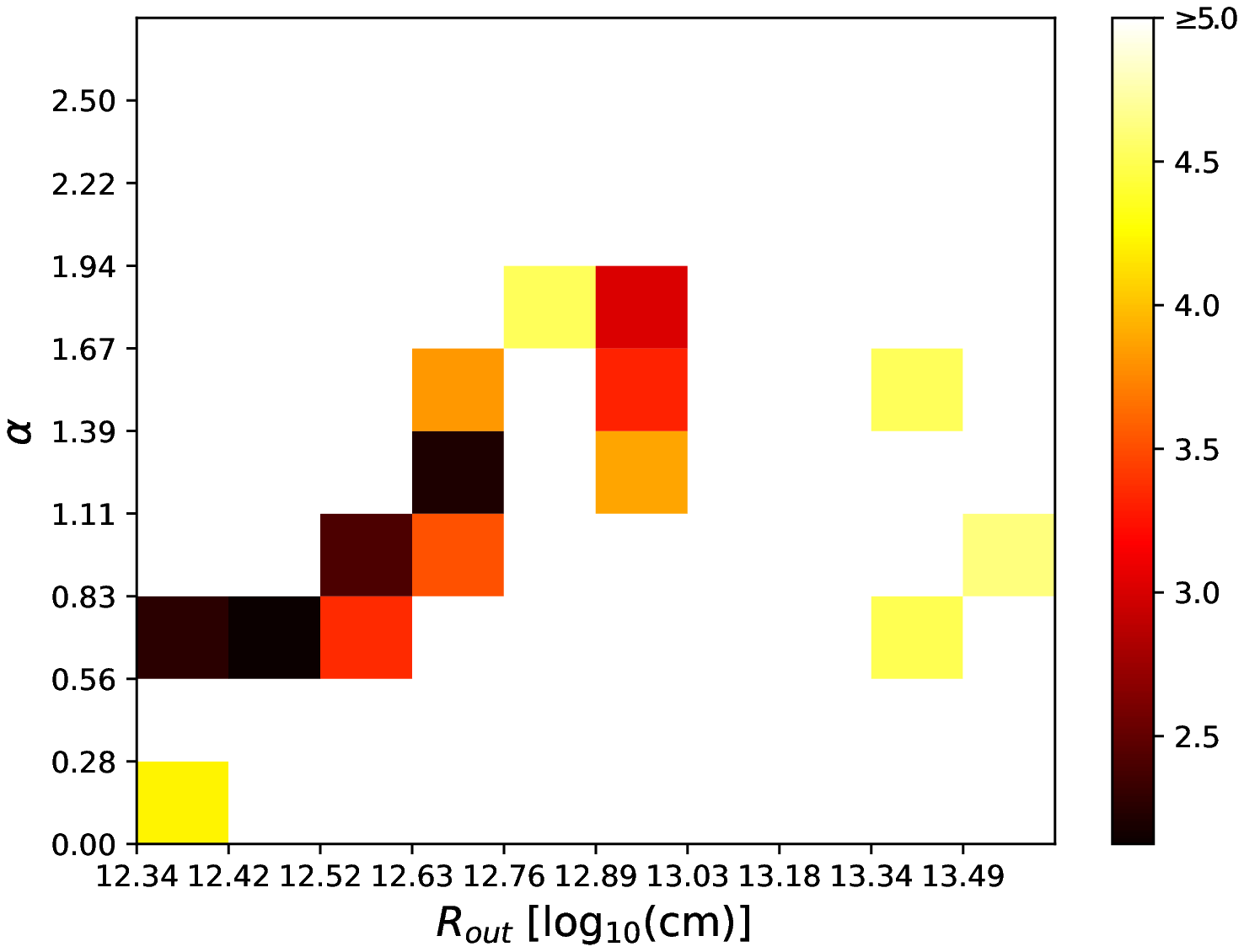}
   \includegraphics[width=6cm]{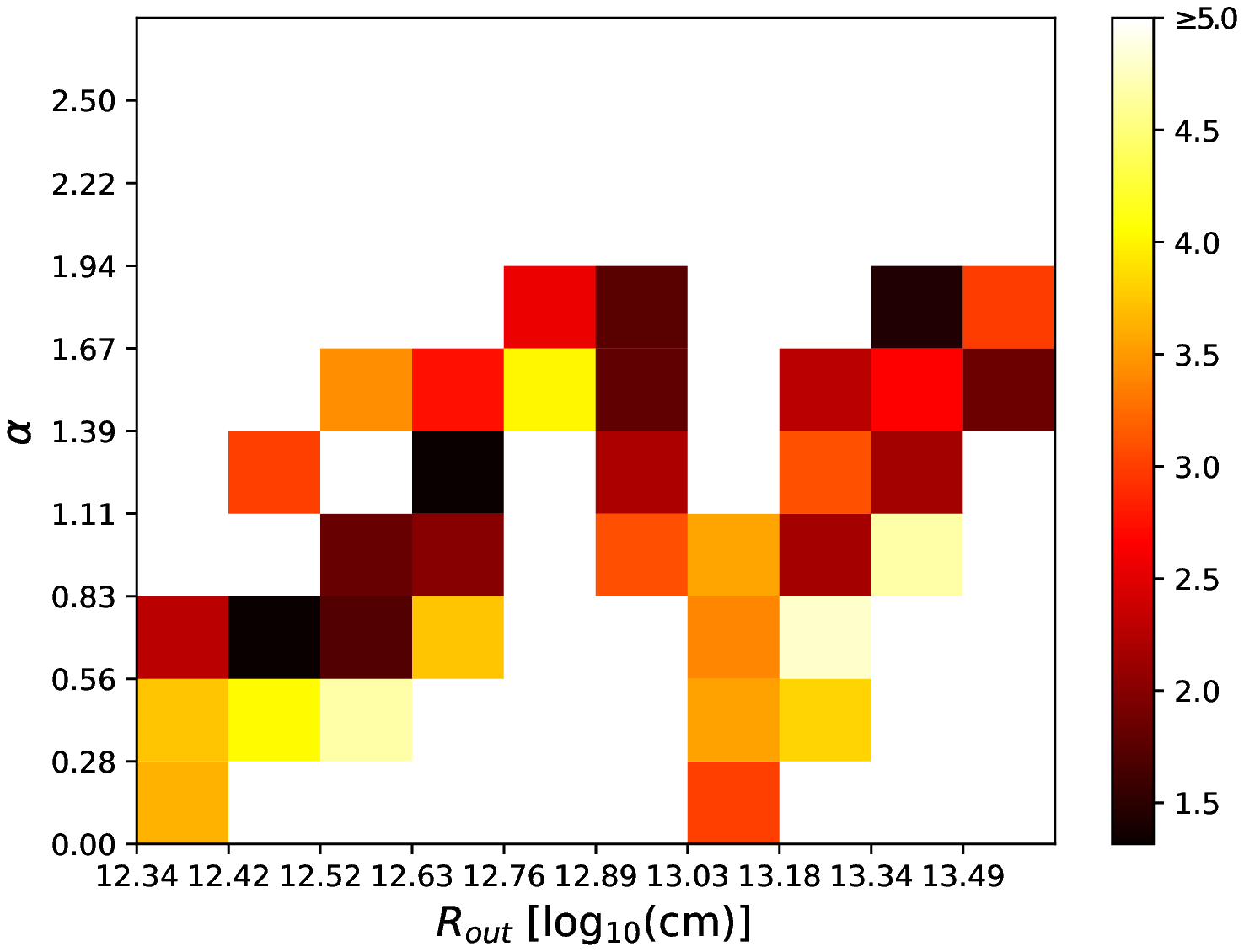}
   \includegraphics[width=6cm]{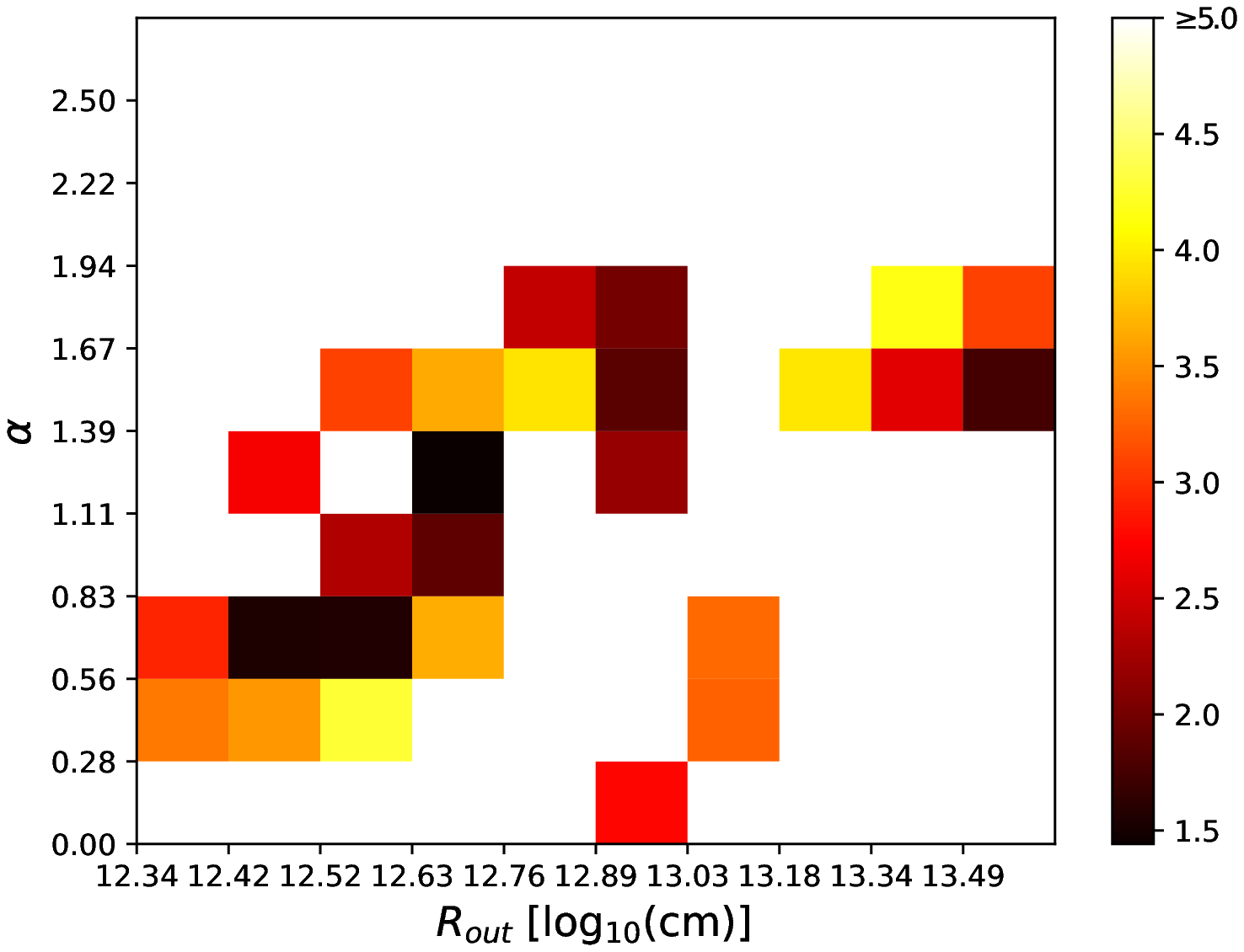}\\

   \includegraphics[width=6cm]{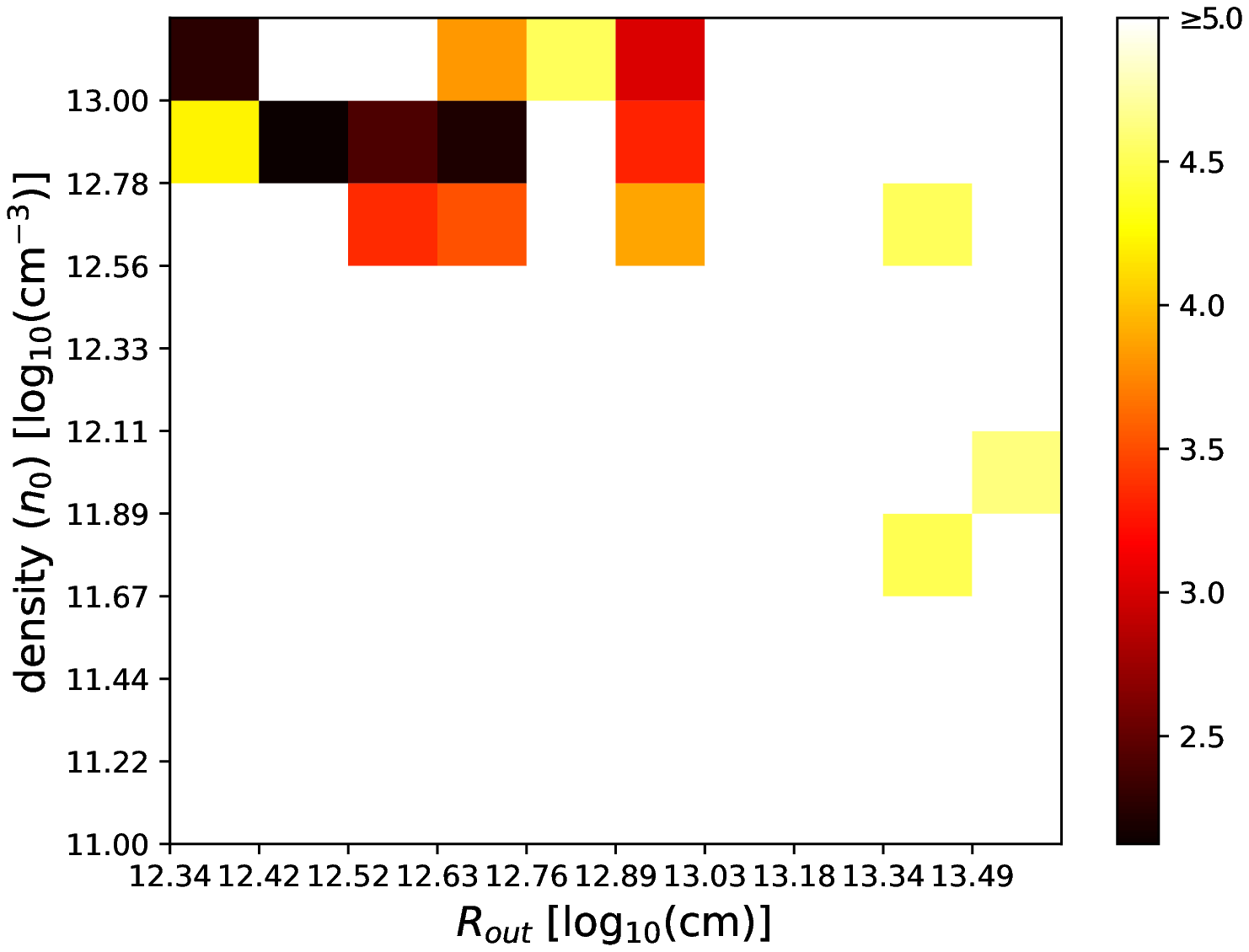}
   \includegraphics[width=6cm]{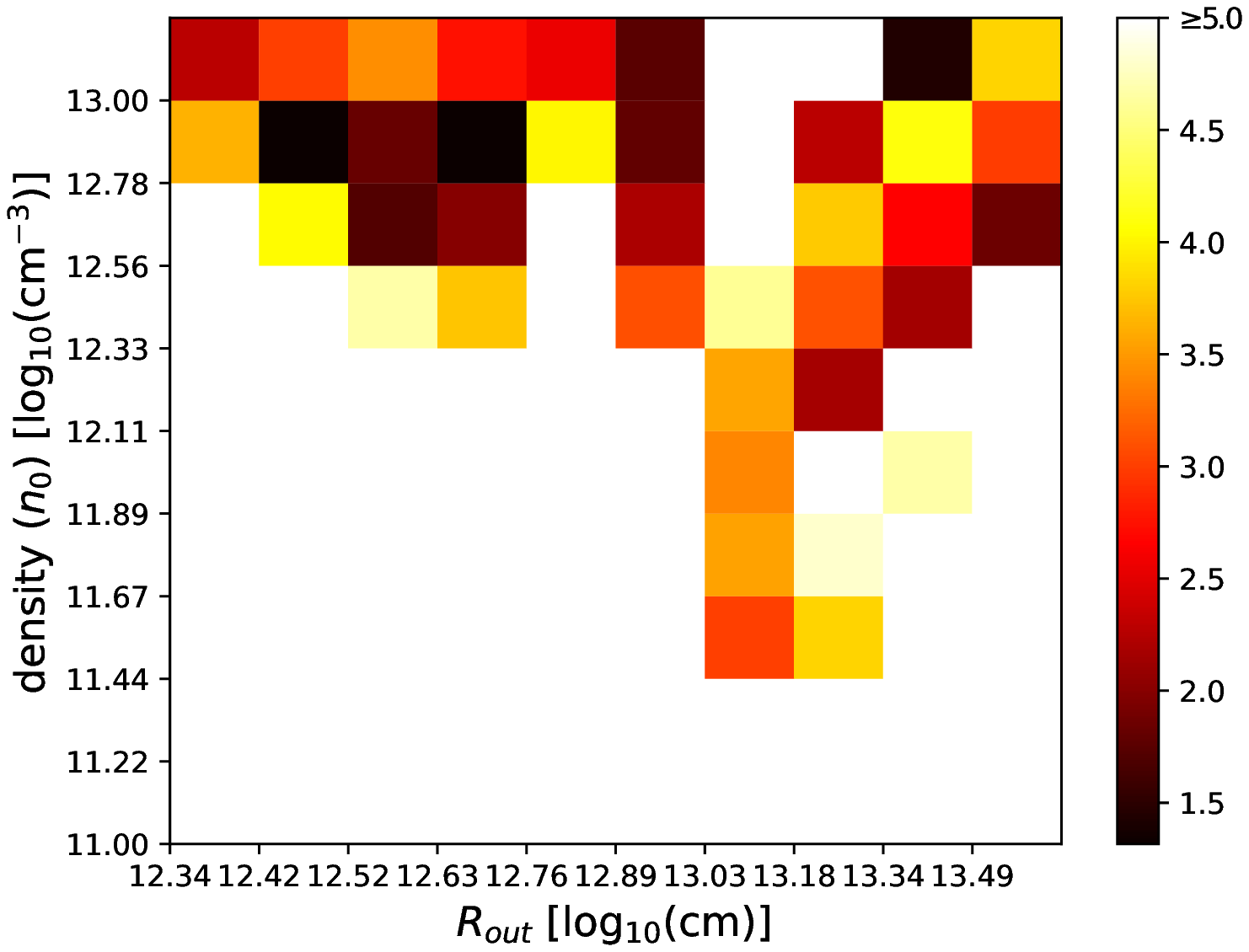}
   \includegraphics[width=6cm]{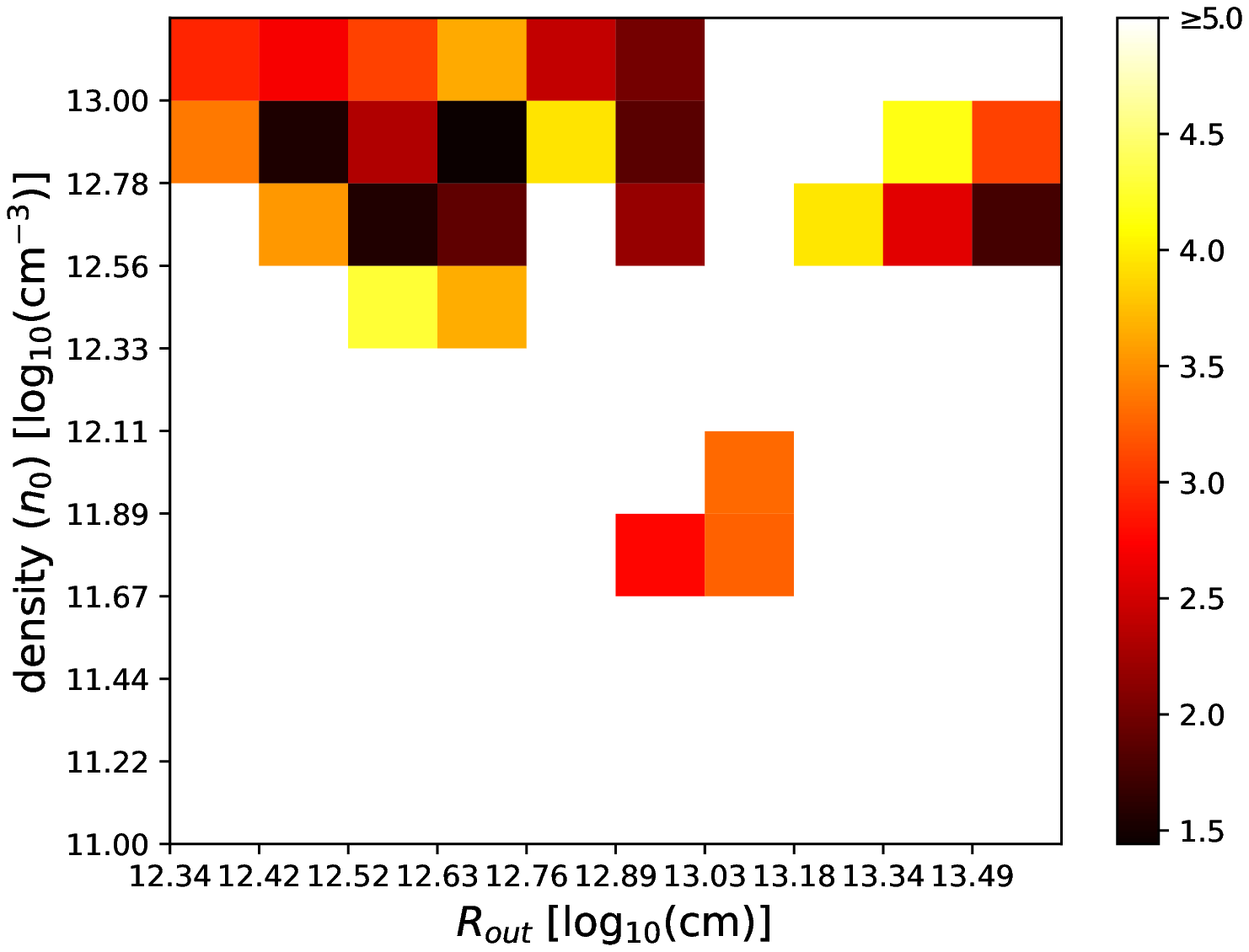}
      \caption{$\chi^2_\nu$ values as a function of the parameters $R_{\rm out}$, $n_0$, and $\alpha$ for three different input spectra.
               In all cases $L_{\rm x}=2.5\times 10^{38}$\,erg,\,s$^{-1}$ (1$-$20\,keV).}
         \label{densprof}
   \end{figure*}

\begin{figure*}
   \centering
   \includegraphics[width=5.7cm]{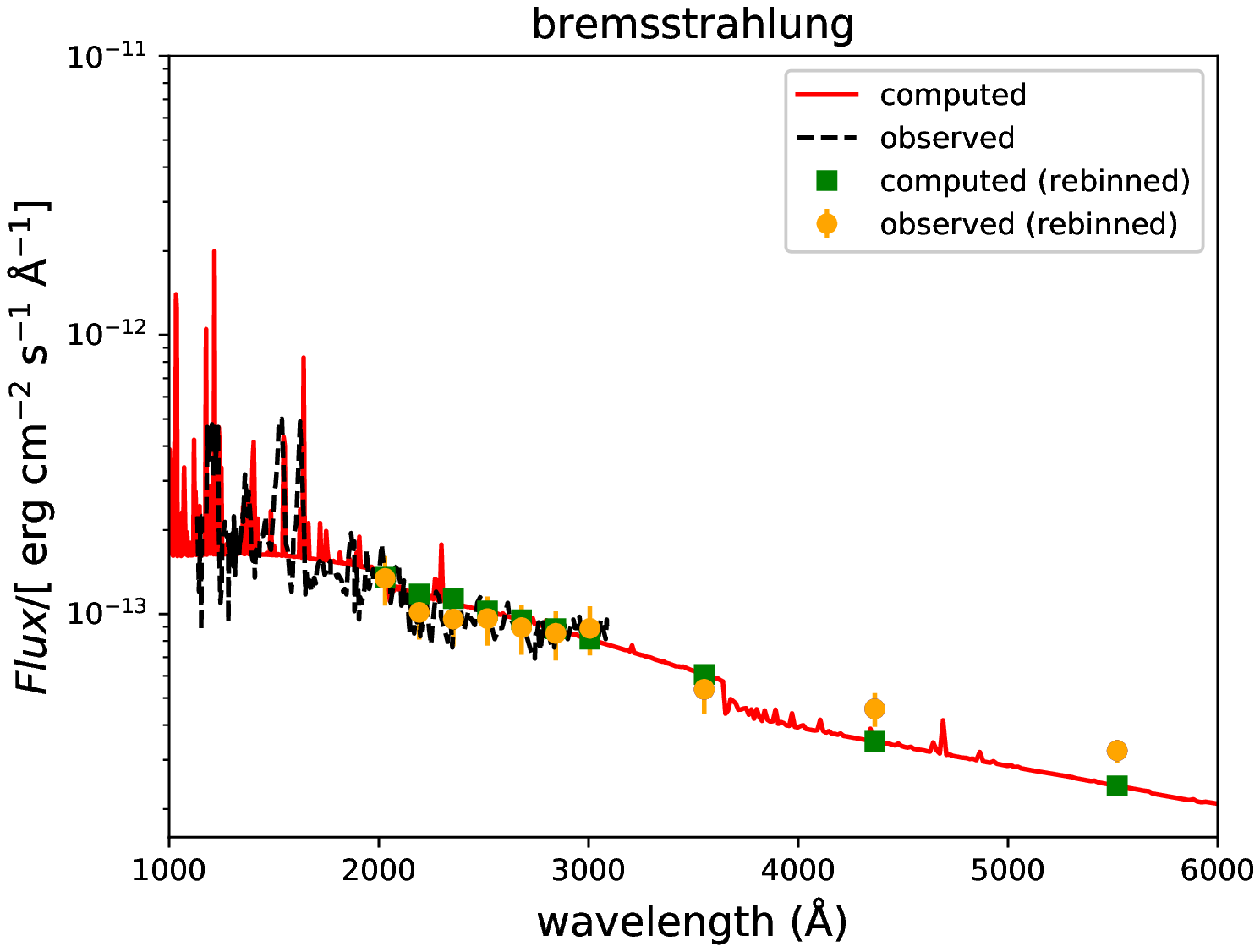}
   \includegraphics[width=5.7cm]{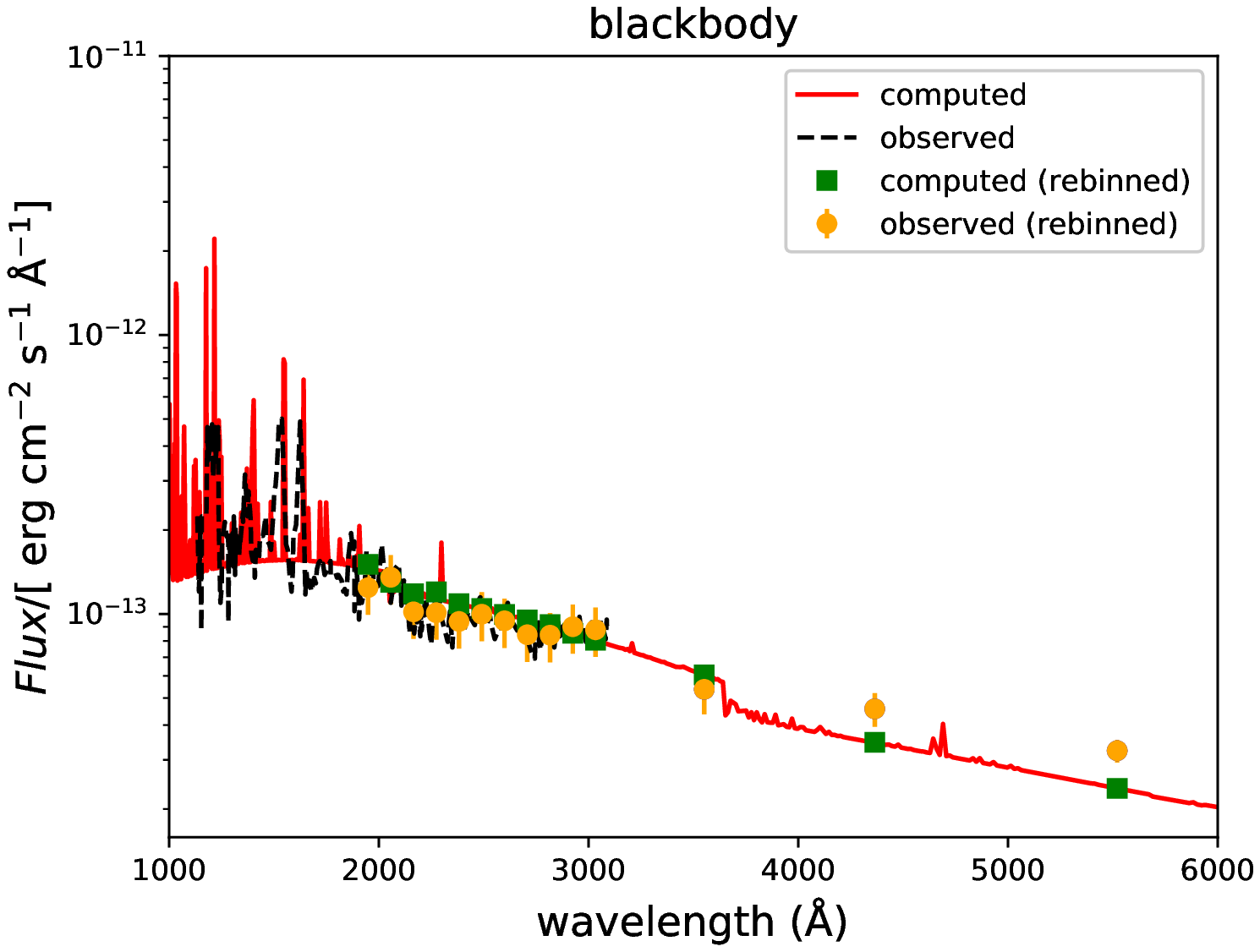}
   \includegraphics[width=5.7cm]{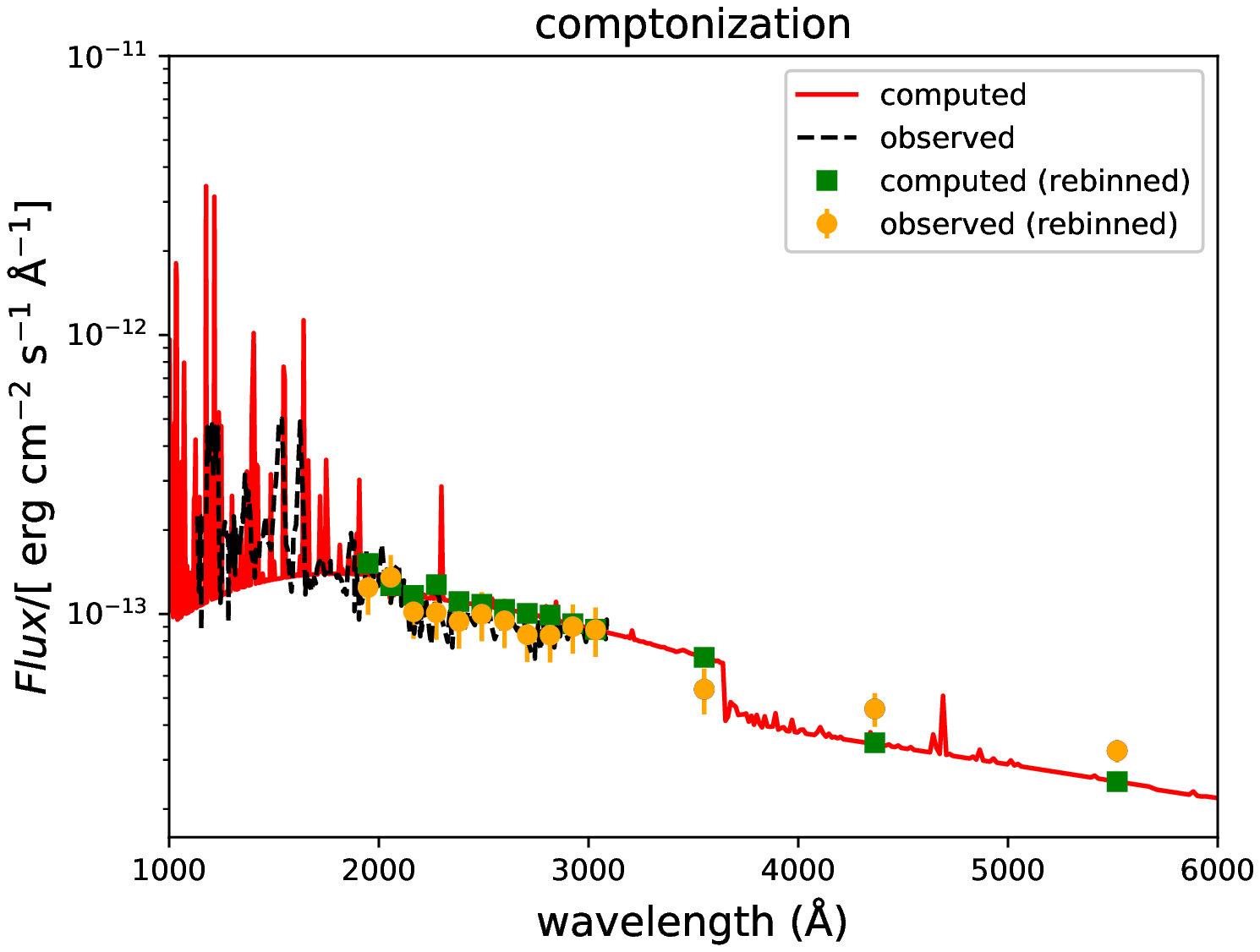} \\
      \caption{\emph{IUE} spectrum and UBV photometric fluxes
      taken during the observation of the outburst of \src\ of 29 April 1981,
      and the best fit spectra computed with CLOUDY assuming the parameters shown in
      Table \ref{Table results1} for three different input spectra.}
         \label{bestfit}
   \end{figure*}

Figure \ref{densprof} shows the $\chi^2_\nu$ values as a function of the fit parameters 
$R_{\rm out}$, $n_0$, and $\alpha$ for each input spectrum (column 1: bremsstrahlung;
column 2: black body; column 3: Comptonization).
Each panel shows the lowest values of the $\chi^2_\nu$ as a function
of two of the three parameters mentioned above.
In all cases, $L_{\rm x}$ is fixed to the best fit value $2.5\times 10^{38}$\,erg\,s$^{-1}$.
Figure \ref{bestfit} shows, for each type of input spectrum,
the computed spectra which best fit the observed spectrum (see Table \ref{Table results1}).
These best fit spectra seem to underestimate
the observed flux at $\lambda \gtrsim 4000$\,\AA. The mismatch for the V magnitude
corresponds to about 2.5\,$\sigma$.
This discrepancy might indicate that the models we assumed for our calculations are too simple.
Further considerations that can be taken into account in future works are the degree of inhomogeneity
of the cloud and the presence of other radiation fields in the system, such as the intrinsic radiation emission
by the donor star, possibly modified by the irradiation flux from the pulsar impinging 
the donor star surface (see Sect. \ref{sect. heating}), and the circumstellar disc
itself, which emits radiation caused by viscosity \citep{Lee91}.


\begin{figure}
   \centering
\includegraphics[bb=85 228 503 533,clip=True,width=8cm]{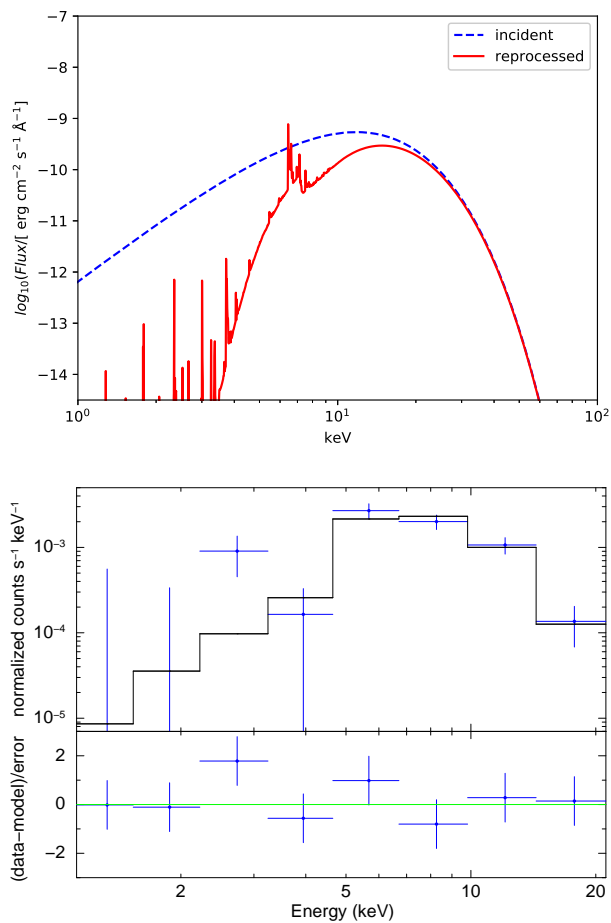}\\
\includegraphics[bb=69 15 583 710,clip=True, angle=-90,width=8cm]{bestfit_simu_bb_mpc.ps}
   \caption{\emph{Top panel:} Comparison between the $kT=2.4$\,keV black-body spectrum
     adopted as input in our {\sc CLOUDY} simulations (blue dashed line)
     and the spectrum emerging from the photoionized circumstellar disc seen edge-on (red solid line).
   \emph{Bottom panel:} Simulated spectrum of \src\ based on the spectrum emerging from
   the nebula, after reprocessing took place (solid red line from top panel),
   for a 4328\,s \emph{Einstein}/MPC observation fitted with an absorbed 
($N_{\rm H} = 9.7{+8.1\atop -3.5} \times 10^{23}$\,cm$^{-2}$) 
black body ($kT=2.2\pm 0.6$\,keV; errors  at 1\,$\sigma$ confidence level).}
         \label{Lx_inc_vs_out}
   \end{figure}

The gas cloud around the system and the X-ray reprocessing caused by it also affect
 the emerging (observable) spectrum in the X-ray band.
As an example, we show in Fig. \ref{Lx_inc_vs_out}
a comparison between the $kT=2.4$\,keV black-body spectrum (Fig. \ref{Lx_inc_vs_out}, top panel, dashed blue line)
used as input in the CLOUDY simulations, and the X-ray spectrum
emerging from the ionized nebula (top panel, solid red line) for the best
fit parameters  in Table \ref{Table results1}.
To check whether the distortion of the emerging spectrum 
could have been detected by the previous observations  
performed by the Monitor Proportional Counter (MPC) on board the \emph{Einstein} satellite,
we simulated the emerging X-ray spectrum of \src\ 
computed with CLOUDY (solid red line in the top panel of the same figure)
using the available \emph{Einstein}/MPC detection redistribution matrices\footnote{\url{ftp://legacy.gsfc.nasa.gov/caldb/data/einstein/mpc/cpf}},
with an exposure time of 4328\,s (see \citealt{Ponman84}).
We obtained a good fit ($\chi^2_\nu=1.04, 5$\,d.o.f.)
assuming an absorbed ($N_{\rm H} = 9.7{+8.1\atop -3.5} \times 10^{23}$\,cm$^{-2}$) 
black body ($kT=2.2\pm 0.6$\,keV; errors  at 1\,$\sigma$ confidence level).
The result is shown in Fig. \ref{Lx_inc_vs_out}.
The output temperature of the black body is only marginally different
from the input temperature,
while the column density is about a factor of five higher than
that measured by \citet{Ponman84},
but we note that the latter is subject to a large uncertainty
due to the limited energy range of the \emph{Einstein}/MPC.

\begin{figure}
   \centering
   \includegraphics[width=\columnwidth]{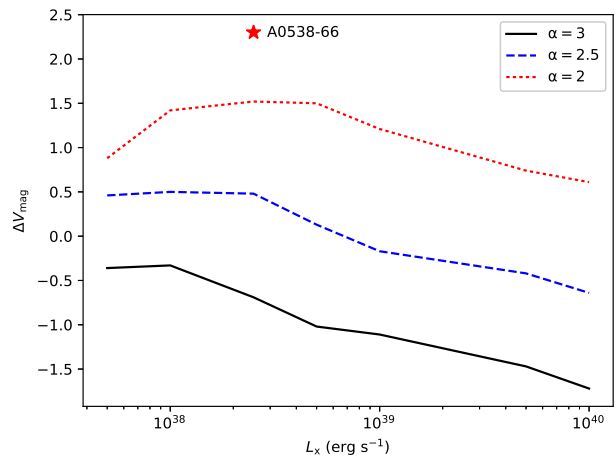}
   \caption{$\Delta V_{\rm mag}$ as a function of the input X-ray luminosity (1--20\,keV) for 
     a black-body input spectrum with $kT=2.4$\,keV and three different slopes 
     of the density profile of the circumstellar disc.
     The red star shows the $\Delta V_{\rm mag}$ of the 29 April 1981 outburst.}
   \label{dischi normali}
   \end{figure}

In the years following the bright optical outbursts studied in this work,
\src\ showed fainter and shorter optical outbursts, while the H$\alpha$
line profiles and the long-term photometric variability indicated the presence
of a circumstellar disc.
To check whether the properties of these weaker optical outbursts
can be explained by a photoionized circumstellar disc
illuminated by an X-ray source,  we assumed
the same orbital and stellar
parameters described in Sect \ref{sect. calculations}.
For the gas cloud, we assumed the typical values reported in \citet{Rivinius13} for the circumstellar discs
in Be/XRBs, namely
$n_0=6\times 10^{12}$\,cm$^{-3}$ ($\rho_0 = n_0 m_{\rm p}= 10^{-11}$\,g\,cm$^{-3}$) 
and $R_{\rm out}=8$\,R$_{\rm d}\approx5.6\times 10^{12}$\,cm.
Typically, $\alpha$ in Be stars is in the range 3--4.
However, in Be/XRBs the circumstellar disc can be truncated by 
tidal interactions with the NS. This leads to a shallower density profile.
Therefore, we assumed the conservative value of $\alpha=3$ found in other 
Be/XRBs \citep{Rivinius13}.
For the input spectrum we used a black body with temperature $kT=2.4$\,keV
and 1$-$20\,keV luminosity ranging from $5\times 10^{37}$\,erg\,s$^{-1}$ to $10^{40}$\,erg\,s$^{-1}$.
We considered the case of a disc seen edge-on in agreement with the results from
\citet{Alcock01} and \citet{McGowan03}.
The results are shown in Fig. \ref{dischi normali} where the difference in magnitude $\Delta V$ 
between the active low state ($V=15$, i.e. we assumed the lowest magnitude from Fig. \ref{figalcock})
and outburst is plotted 
for each input X-ray luminosity.
Also shown in the  figure are two unusual  scenarios for Be/XRBs,
having density profiles with extreme slopes, $\alpha=2.5$ and $\alpha=2$.
For comparison, we  plotted $\Delta V$ observed
on 29 April 1981 \citep{Charles83}.
Figure \ref{dischi normali} shows that for a
density profile of $\alpha = 3$, the increase in the $V$ magnitude
during the outburst is negligible compared to the magnitude of the system
star plus circumstellar disc observed during the active-low state.
This reinforces our previous findings that the gas cloud
around the system on 29 April 1981 was not a classical circumstellar disc and,
albeit indirectly, supports the lack of bright optical outbursts
associated with the X-ray outbursts in the other Be/XRBs.
For $\alpha=2$ and $\alpha=2.5$ there is an increase of $\Delta V \approx 0.5 - 1.5$,
similar to that seen in the low-luminosity optical outbursts displayed
by \src\ (see Fig. \ref{figalcock} and \citealt{Ducci16}).
Although these steep density profiles are not observed in Be/XRBs,
it is important to note that a value of $\alpha=2.1$ was measured by \citet{Waters88} in \src\ during
an infrared observation on 9 October 1981 carried out during an outburst
not observed in the optical bands. Nonetheless, \citet{Waters88}
were able to extrapolate an expected $V$ magnitude during the outburst
of $V\approx 13.6$, which is in good agreement with the value at the peak
of the red line ($\alpha=2$) in Fig. \ref{dischi normali}.

%

\subsection{X-ray irradiation of the surface of the donor star}
\label{sect. heating}

\begin{figure}
   \centering
   \includegraphics[width=\columnwidth]{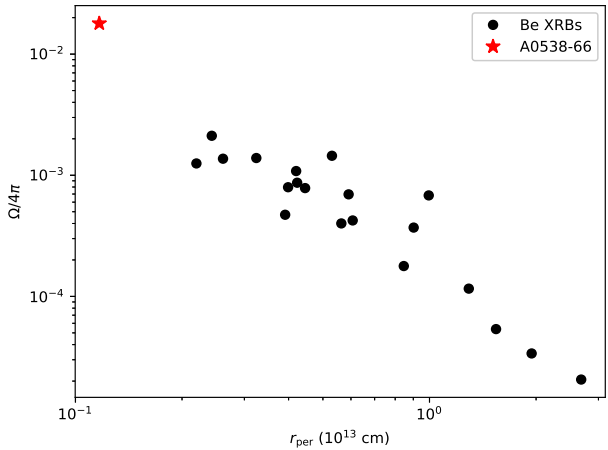}
   \caption{Fraction of solid angle of the X-ray radiation emitted 
     by the NS and intercepted by the donor star at periastron,
     for different Be/XRBs with known orbital parameters, compared to \src.}
   \label{solid angle}
   \end{figure}

In the previous section, we considered the case of 
optical outbursts powered by the reprocessing of X-ray radiation in
the envelope surrounding the binary system.
However, as already noticed by \citet{Charles83}, other mechanisms could cause the optical outbursts.
Here we consider the effects of the irradiation of the surface of the Be star by X-ray photons
emitted by the accreting NS.
Because of the high eccentricity and small orbital period
of \src, the separation between the two stars at periastron is such that the fraction 
of X-ray radiation emitted by the NS (assuming isotropic emission) intercepted
by the companion star is higher than in the other Be/XRBs with known orbital parameters.
This characteristic can be seen in Fig. \ref{solid angle}, where we show---for a sample of Be/XRBs with known system parameters
(orbital period, eccentricity, mass, and radius of the donor star or its spectral type)---the fraction of NS radiation intercepted by the donor star at periastron as a function of the 
separation of the stars at periastron.
We considered for this plot  the binary systems with OBe stars reported in the catalogue of
\citet{Liu06} and Table 2 of \citet{Klus14} with known orbital period and eccentricity,
spectral type, and luminosity class of the donor star.
In some cases the masses and radii of the donor stars were reported in \citet{Okazaki01}.
In the other cases, we derived them from the spectral and luminosity class
using the Catalogue of Apparent Diameters and Absolute Radii of Stars (CADARS)  \citep{Pasinetti01} and the catalogue of \citet{Hohle10}.

\begin{figure}
   \centering
\includegraphics[width=\columnwidth]{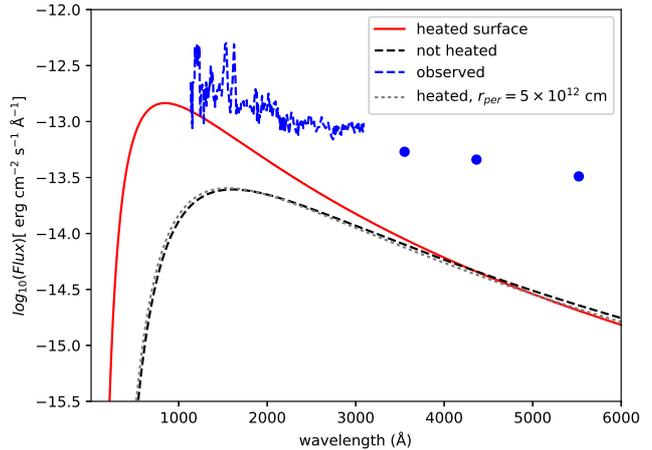}
   \caption{Comparison between the optical/UV spectrum of the donor star of \src\
     not irradiated by the NS  and the heated spectrum for the irradiated surface of the donor star.
     We assumed here $\eta=0.5$ and an input X-ray luminosity of $L_{\rm x}=10^{39}$\,erg\,s$^{-1}$.}
   \label{heating spectra comparison}
   \end{figure}

We calculated the expected optical/UV spectrum 
emitted by the surface of the donor star irradiated by
the NS at the periastron passage using a model developed to study these effects 
in binary systems that hosts a bright accreting X-ray source
(e.g. \citealt{Beech85} and \citealt{Tjemkes86}).
In this model, the X-rays emitted by the compact object are absorbed by the atmosphere
of the companion star and then reradiated at lower energy. 
For each surface element of the Be star, the local effective temperature
increases according to
\begin{equation} \label{eq. heating}
T_{\rm i}^4 = T_{\rm e,i}^4 + (1-\eta) \frac{L_{\rm x} \cos \phi}{4 \pi \sigma_{\rm SB} r^2} \mbox{ ,}
\end{equation}
where $T_{\rm e,i}$ is the effective temperature of the surface element in the
absence of X-ray irradiation, $\eta$ is the fraction of reflected X-rays (albedo),
$L_{\rm x}$ is the luminosity of the X-ray source (assumed to be emitted
isotropically from the NS), $\sigma_{\rm SB}$ is the Stefan-Boltzmann constant,
$\phi$ is the angle between the normal to the surface element and the direction to the X-ray source, and
$r$ is the distance from the X-ray source to the surface element.
In our calculations, we neglect gravity darkening effects and
we assume a spherically symmetric donor star and $\eta =0.5$.
The same value for the albedo was assumed by \citet{Charles83}
to calculate the effects of X-ray heating on the Be star of \src.
In our calculations, the total spectrum and luminosity of the heated side of the companion is then
determined by the sum of the black-body spectra emitted by each surface element. 
Figure \ref{heating spectra comparison} shows the spectrum irradiated (red line) by the heated side of the donor
star during the periastron passage of the NS emitting at $L_{\rm x}=10^{39}$\,erg\,s$^{-1}$
compared to the undisturbed spectrum of the donor star ($T_{\rm eff}=18000$\,K, $L_{\rm opt}=4.7\times 10^{37}$\,erg\,s$^{-1}$,
from \citealt{Maraschi83}) and the \emph{IUE}+\emph{UBV} spectrum.
We also show in the same figure that
the X-ray heating would be negligible if the NS were
located at a greater distance, for example, $r_{\rm per}=5\times 10^{12}$\,cm,
which is the average periastron separation of the Be/NS binaries (dotted grey line).
In conclusion, the X-ray heating of the surface of the donor star
of \src\ is non-negligible, but it is not enough to explain alone
 the \emph{IUE}+\emph{UBV} spectrum observed during the bright optical outburst of 29 April 1981.

\subsection{X-ray heating of the surface of an accretion disc}
\label{sect irr accr disc}

\begin{figure}
   \centering
   \includegraphics[width=\columnwidth]{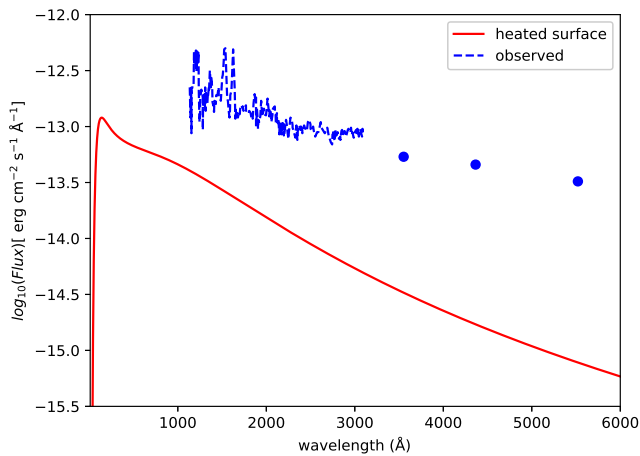}
   \caption{Comparison between the observed optical/UV spectrum of \src\ during an outburst and
           the calculated spectrum from an irradiated accretion disc.}
   \label{heating spectra comparison2}
   \end{figure}

\citet{Charles83} ruled out the possibility that the UV and optical emission of \src\
during the outburst was produced by a Shakura-Sunyaev disc, for which the known relation $F_\nu \propto \nu^{1/3}$
is expected.
However, the spectral shape of an accretion disc irradiated by the X-ray photons from the 
accreting NS is significantly different from $F_\nu \propto \nu^{1/3}$ (see e.g. the review by \citealt{Hynes10}).
Although it is not clear whether an accretion disc was actually present
during the optical outburst studied in this work,
it has recently been argued that 
an accretion disc might be present during the most recent periastron passages \citep{Rajo17}.
It is therefore worth examining this case.
We followed the calculations proposed by \citet{Vrtilek90} to model
the spectrum of an irradiated disc.
In their model, the temperature of the irradiated disc at a radius $r$ is given by
\begin{equation} \label{eq vrtilek}
T(r) = \left [ \frac{f L_x \sqrt{kGM_{\rm NS}/(\mu m_p)}}{14 \pi \sigma_{\rm SB} G M_{\rm NS}} \right ]^{2/7} \mbox{ ,}
\end{equation}
where $f=0.5$ is the fraction of the X-ray radiation absorbed by the disc surface, $k$ is the Boltzmann constant,
$\mu=0.5$ is the mean molecular weight, $m_p$ is the proton mass, $L_x$ is the X-ray luminosity of the NS,
$G$ is the gravitational constant, and $M_{\rm NS}$ is the mass of the NS.
We assume that the spectrum of the ring of the accretion disc at radius $r$ is described by 
a black body with temperature given by Eq. \ref{eq vrtilek}.
The spectrum of the entire accretion disc is then given by the sum of all the ring spectra.
We calculated the emission from an irradiated disc assuming different values of the input X-ray luminosity,
without finding an appropriate solution.
As an example, Fig. \ref{heating spectra comparison2} shows
a comparison between the observed UV and optical emission of \src\ during 
the outburst and the calculated continuum emission of an accretion disc with
size $R_{\rm disc} = 3\times 10^{11}$\,cm irradiated by a NS emitting at $L_{\rm x}=10^{39}$\,erg\,s$^{-1}$.

\section{Conclusions}

We examined different mechanisms to explain the bright optical outbursts
shown by \src.
We found, through CLOUDY simulations, that the X-ray photoionization
of a spherical shell around the binary system can explain reasonably well the
observed \emph{IUE} spectrum and the \emph{UBV} magnitudes taken during the outburst of 29 April 1981.
We found the best fit between the computed and observed spectra
by assuming for the input X-ray spectrum a black body with temperature of $kT=2.4$\,keV
and a $1-20$\,keV luminosity of $L_{\rm x}=2.5\times 10^{38}$\,erg\,s$^{-1}$
(i.e. the spectrum observed by \citet{Ponman84} during another outburst that
reached similar X-ray luminosities).
On the basis of our simulations, the properties of the gas cloud 
of \src\ during the bright optical outbursts seem consistent
with the qualitative scenario proposed by
\citet{Densham83}, \citet{Howarth84}, and \citet{Maraschi83}, where
a spherical gas cloud forms around the binary system
from the material tidally displaced by the NS from the outer layers of the donor star
over many orbits.
The bright optical outbursts are then powered by the X-ray photons produced by the accreting 
NS, which are reprocessed in the envelope.
We found that our simulations can explain the density profile and $V$ magnitude measured
by \citet{Waters88} during the outburst of 9 October 1981.
Nonetheless, there are some points in our results that need to be  clarified:
\begin{itemize}
\item the column density value obtained from the fit of
  the X-ray spectrum (1--20\,keV) emerging from the photoionized nebula
  with an absorbed black body (see Sect. \ref{reprocessing} and Fig. \ref{Lx_inc_vs_out})
  is about five times larger than the highest value measured by \citet{Ponman84}.
  However, \citet{Ponman84} noted that the column density of \src\ was highly variable during the outburst.
  In addition, we note that the $\chi^2_\nu=1.633$ (3 d.o.f.) reported in Table 1 of \citet{Ponman84} 
  shows that the assumed fitting model (absorbed black body) does not describe well the \emph{Einstein}/MPC data,
  and the uncertainties on the spectral parameters were not reported; 

\item the luminosities of the UV lines \ion{C}{IV}\,$\lambda$1550 and \ion{He}{II}\,$\lambda$1640
  reported in Table \ref{Table results1} are, within an order of magnitude, in agreement with those observed; however, their luminosity is systematically lower than that observed;
\item the best fit spectra shown in Fig. \ref{bestfit} seem to underestimate the observed flux at $\lambda \gtrsim 4000$\,\AA.
\end{itemize}
These issues could indicate either that the photoionization of a cloud surrounding the
system is not the correct explanation for the observed optical outbursts,
or that our model is incomplete.
In this regard, we have shown that another peculiarity of \src\ with respect to the other Be/XRBs
is that the radiation produced by the X-ray heating of the surface of the donor
star is non-negligible during the bright X-ray outbursts (Sect. \ref{sect. heating}).
Although this mechanism cannot explain alone the observed UV/optical
emission during the bright outbursts, we cannot exclude, for example,
that the reprocessing in the gas cloud around the system of the
radiation emitted by the heated surface of the donor
may account for the missing optical emission mentioned above.
Future investigations based on more complex models will help to explain
the cause of the small discrepancies with the observations that we have reported here.

\begin{acknowledgements}
We thank the anonymous referee for the useful comments that helped to improve the paper.
We acknowledge support from the High Performance and 
Cloud Computing Group at the Zentrum f\"ur Datenverarbeitung of the 
University of T\"ubingen, the state of Baden-W\"urttemberg through bwHPC,
and the German Research Foundation (DFG) through grant No. INST 37/935-1 FUGG.
SM and PR acknowledge financial contribution from the agreement ASI-INAF n.2017-14-H.0.
PR acknowledges contract ASI-INAF I/004/11/0. 
KH is grateful to the Polish National
Science Center for support under grant No.
2015/17/B/ST9/03422.
\end{acknowledgements}

\bibliographystyle{aa} 
\bibliography{lducci_1col}

\end{document}